\documentclass[twocolumn,showpacs,preprintnumbers,amsmath,amssymb]{revtex4}
\usepackage{graphicx}
\usepackage{dcolumn}
\usepackage{bm}

\begin{document}
\title{Searches for the technicolor signatures via $gg \rightarrow W^{\pm}\pi_t^{\mp}$ at the Large Hadron Collider}
\author{Jinshu Huang}
\email{jshuang@foxmail.com} \affiliation{College of Physics $\&$ Electronic Engineering, Nanyang Normal University, Nanyang 473061,
People's Republic of China; \\ College of Physics $\&$ Information Engineering, Henan Normal University, Xinxiang 453007, People's
Republic of China}
\author{Taiping Song}
\author{Shuaiwei Wang}
\affiliation{College of Physics $\&$ Electronic Engineering, Nanyang Normal University, Nanyang 473061, People's Republic of China}
\author{Gongru Lu}
\email{lugongru@sina.com} \affiliation{College of Physics $\&$ Information Engineering, Henan Normal University, Xinxiang 453007,
People's Republic of China}

\date{\today}

\begin{abstract}
In this paper we calculate the production of a charged top pion in association with a $W$ boson via $gg$ fusion at the CERN's Large Hadron Collider in the context of the topcolor assisted technicolor model. We find that, the total cross section of $pp \rightarrow gg \rightarrow W^{\pm}\pi_t^{\mp}$, is several dozen femtobarns with reasonable values of the parameters, and the total cross section of $pp \rightarrow W^{\pm}\pi_t^{\mp}$ can reach a few hundred femtobarns when we consider the sum of the contributions of these two parton subprocesses $gg \rightarrow W^{\pm}\pi_t^{\mp}$ and $b\bar{b} \rightarrow W^{\pm}\pi_t^{\mp}$.
\end{abstract}


\maketitle

\section{\label{sec:level1}Introduction}

The search for Higgs bosons and new physics particles and the study of their properties are among the prime objectives of the large hadron collider (LHC) \cite{Quigg2007}. In the Standard Model (SM), a single neutral Higgs boson is predicted as a direct consequence of the mechanism of electroweak symmetry breaking (EWSB). Moreover, charged Higgs bosons are predicted in extended versions of the SM, such as the minimal supersymmetric standard model (MSSM). Since the discovery of such an additional Higgs boson will be the evidence of new physics beyond the SM, there is increasing interest in theoretical and experimental studies to provide the basis for its accurate exploration.

Recently, lots of studies of the neutral or charged Higgs production at the LHC have been finished \cite{Jakobs2009}. For the production of charged Higgs boson in association with a $W$ boson in the MSSM, Ref. \cite{Dicus1989} investigates $b\bar{b} \rightarrow W^{\pm}H^{\mp}$ at the tree level and $gg \rightarrow
W^{\pm}H^{\mp}$ at one loop. The electroweak corrections and QCD corrections to $b\bar{b} \rightarrow W^{\pm}H^{\mp}$ have already been calculated in Ref. \cite{Yang2000}, which shows that a favorable scenario for $W^{\pm} H^{\mp}$ associated production would be characterized by the conditions that $m_H >m_t-m_b$ and that $\tan\beta$ are either close to unity or of order $m_t/m_b$, then the $H^{\pm}$ bosons could not spring from on-shell top quarks and could be copiously produced at hadron colliders. A complete one-loop calculation of the loop-induced subprocess $gg \rightarrow W^{\pm}H^{\mp}$ is presented in the framework of the MSSM \cite{Brein2001}, which shows the cross section of $gg \rightarrow W^{\pm}H^{\mp}$ can be comparable to that of $b\bar{b} \rightarrow W^{\pm}H^{\mp}$ due to the large number of gluons in the high energy proton beams at the LHC.

Technicolor theory \cite{Weinberg1976} is one of the important candidates for probing new physics beyond the SM, especially the topcolor assisted technicolor (TC2) model proposed by C. T. Hill \cite{Hill1995} --- this combines technicolor with topcolor, with the former being mainly responsible for EWSB and the latter for generating a major part of the top quark mass. If technicolor is actually responsible for EWSB, there are strong phenomenological arguments that its energy scale is at most a few hundred GeV and that the lightest technicolor pions are within reach of the ATLAS and CMS experiments at the LHC \cite{Carena2008}. The TC2 model predicts three top pions ($\pi^0_t, \pi^{\pm}_t$), one top Higgs ($h^0_t$) and the new gauge bosons ($Z', B$) with large Yukawa couplings to the third generation quarks, so these new particles can be regarded as a typical feature of this model. Lots of signals of this model have already been studied in the work environment of linear colliders and hadron-hadron colliders \cite{Wang2002}, but most attention has been focused on the neutral top pion and new gauge bosons. Here we wish to discuss the prospects of charged top pions.

For the production of charged top pion in association with a $W$ boson at the LHC, there are mainly two partonic subprocesses that contribute to the hadronic cross section $pp \rightarrow W^{\pm}\pi^{\mp}_t$: the $b\bar{b}$ annihilation and the $gg$ fusion. Ref. \cite{Huang2004} has already studied the process of
$W^{\pm}\pi^{\mp}_t$ associated production via $b\bar{b}$ annihilation at the tree level and the one-loop, which shows that the total cross section $\sigma(p\bar{p} \rightarrow b\bar{b} \rightarrow W^{\pm} \pi^{\mp}_t)$ is rather large when $\pi^{\pm}_t$ is not very heavy. In this paper we shall discuss the production of top pions $\pi_t^{\pm}$ in association with SM gauge bosons $W^{\mp}$ via $gg$ fusuion, including the contributions arising from top pions $\pi^0_t, \pi^{\pm}_t$ and top Higgs $h^0_t$, to search for new physics particles and test the TC2 model.

\section{\label{sec:level2} The partonic process $gg \rightarrow W^-\pi^+_t$}

\begin{figure}

\ \

\vspace{-3.5cm}

\includegraphics{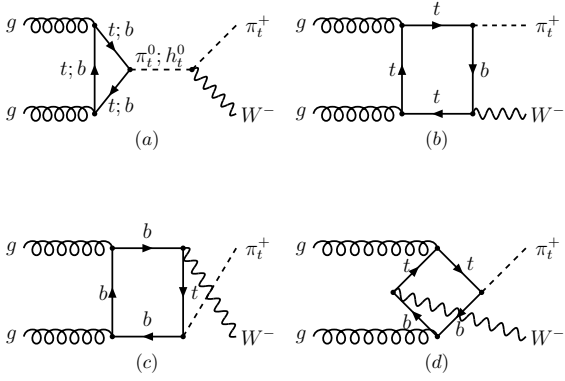}

\vspace{-20cm}

\caption{\label{fig:eps1} Feynman diagrams for the technicolor correction to the $gg \rightarrow W^-\pi_t^+$ process.}
\end{figure}

In the TC2 model, there is no tree-level contribution to the subprocess $gg \rightarrow W^-\pi^+_t$. This process is induced at the one-loop level by diagrams with quark-loops. The quark-loop diagrams can be subdivided into box-type diagrams and into triangle diagrams. In each box diagram, the quarks in the loop couple directly to the outgoing charged Higgs boson $\pi^+_t$, while in the triangle diagrams the quarks couple to one of the neutral top pion and top Higgs boson. The Feynman diagrams of the process $gg \rightarrow W^{\pm} \pi^{\mp}_t$ are shown in Fig. \ref{fig:eps1}. The relevant Feynman rules are given in Refs.
\cite{Hill1995,Kaul1983}.

In our kinematical conventions, the momenta of the initial state gluons, $p_1$ and $p_2$, are chosen as incoming and outgoing for the momenta, $k_1$ and $k_2$, of the final state particles
\begin{eqnarray}
g(p_1,a,\sigma_1)+g(p_2,b,\sigma_2) \rightarrow W^-(k_1,\lambda)+\pi_t^+(k_2),
\end{eqnarray}

Besides being characterized by their momenta, the initial state gluons are characterized by their color indices $a$, $b$ and their helicities $\sigma_1$,
$\sigma_2$($=\pm 1$), and the final state $W$ boson is characterized by its helicity $\lambda$ ($=0, \pm 1$).

The Mandelstam variables are defined as
\begin{eqnarray}
\hat{s}=(p_1+p_2)^2=(k_1+k_2)^2, \nonumber \\ \hat{t}=(p_1-k_1)^2=(p_2-k_2)^2, \nonumber \\ \hat{u}=(p_1-k_2)^2=(p_2-k_1)^2.
\end{eqnarray}
with obeying
\begin{eqnarray}
\hat{s}+\hat{t}+\hat{u}=m^2_W+m^2_{\pi_t}.
\end{eqnarray}

Using the relation
\begin{eqnarray}
\sum^8_{a,b=1} \Big [ {\rm Tr}\Big(\frac{\lambda^a}{2}\frac{\lambda^b}{2}\Big ) \Big ]^2= 2,
\end{eqnarray}
we can write the spin- and color-averaged cross section for the parton process
\begin{eqnarray}
\frac{{\rm d} \hat{\sigma}}{{\rm d} t} =\frac{1}{16 \pi \hat{s}^2} \sum_{\lambda=0,\pm 1} \frac{1}{4} \sum_{\sigma_1,\sigma_2=\pm 1}
\frac{1}{32} \big |M_{\sigma_1 \sigma_2 \lambda} \big |^2,
\end{eqnarray}
which contains the helicity amplitudes
\begin{eqnarray}
M_{\sigma_1 \sigma_2
\lambda}=\varepsilon^{\mu}_{\sigma_1}(p_1)\varepsilon^{\nu}_{\sigma_2}(p_2) \varepsilon^{\rho}_{\lambda}(k_1)\tilde{M}_{\mu\nu\rho},
\end{eqnarray}
where $\varepsilon^{\mu}_{\sigma_1}(p_1)$, $\varepsilon^{\nu}_{\sigma_2}(p_2)$, $\varepsilon^{\rho}_{\lambda}(k_1)$ are the polarization vectors for
the incoming gluons and outgoing $W$ bosons, respectively.

In the parton center-of-mass (CM) frame, the momenta may be expressed by
\begin{eqnarray}
p_1^{\mu} = \bigg ( \frac{\sqrt{\hat{s}}}{2},0,0,\frac{\sqrt{\hat{s}}}{2} \bigg ),
p_2^{\mu} = \bigg ( \frac{\sqrt{\hat{s}}}{2},0,0,-\frac{\sqrt{\hat{s}}}{2} \bigg ), \\
k_1^{\mu} = \bigg ( E_W,\vec{p}_W \bigg ) = \bigg ( E_W, |\vec{p}_W|\sin\theta,0,|\vec{p}_W|\cos\theta \bigg ),
\end{eqnarray}
and then the polarization vectors are given by
\begin{eqnarray}
\varepsilon^{\mu}_{\sigma_1}(p_1) = \frac{1}{\sqrt{2}} \bigg (0, 1, i \sigma_1, 0 \bigg ),\\
\varepsilon^{\mu}_{\sigma_2}(p_2) =  \frac{1}{\sqrt{2}} \bigg (0, 1, -i \sigma_1, 0 \bigg ),
\end{eqnarray}
 and
\begin{eqnarray}
\varepsilon^{\mu}_{\lambda=0}(k_1) = \bigg (\frac{|\vec{p}_W|}{m_W},\frac{E_W}{m_W}\sin\theta,0, \frac{E_W}{m_W}\cos\theta \bigg ),\\
\varepsilon^{\mu}_{\lambda=\pm 1}(k_1) = \frac{1}{\sqrt{2}} \bigg (0, i\lambda \cos \theta, 1, -i \lambda \sin\theta \bigg ),
\end{eqnarray}

The remaining tensor $\tilde{M}_{\mu\nu\rho}$ can be expressed in terms of three- and four- point scalar integrals \cite{Passarino1979}, and their analytical expressions are tedious, so we do not present them.

Finally, the integrated partonic cross section for the process $gg \rightarrow W^{\pm}\pi_{t}^{\mp}$ is
\begin{eqnarray}
\hat{\sigma}(\hat{s})=\int^{\hat{t}_{+}}_{\hat{t}_{-}}{\rm d} \hat{t} \frac{{\rm d} \hat{\sigma}}{{\rm d} \hat{t}},
\end{eqnarray}
with
\begin{eqnarray}
\hat{t}_{\pm} &=& \frac{m_W^2+m_{\pi_t}^2-\hat{s}}{2} \nonumber \\
&& \pm\frac{1}{2} \sqrt{ \left [\hat{s}-(m_W+m_{\pi_t})^2 \right ]
\left [\hat{s}-(m_W-m_{\pi_t})^2 \right ]}.  \nonumber \\ && \
\end{eqnarray}
The total hadronic cross section for $pp \rightarrow gg \rightarrow W^{\pm}\pi_t^{\mp}$  can be obtained by folding the subprocess cross section $\hat{\sigma}$ with the parton luminosity \cite{Dicus1989}
\begin{equation}
\sigma(s)=\int_{(m_W+m_{\pi_t})/\sqrt{s}}^1 {\rm d}z \frac{{\rm d}L}{{\rm d} z}\hat{\sigma}(gg \rightarrow W^{\pm}\pi_t^{\mp}\ {\rm at}\ \hat{s}=z^2 s).
\end{equation}
Here  $\sqrt{s}$ and $\sqrt{\hat{s}}$ are the CM energies of the $pp$ and $gg$ states, respectively, and ${\rm d}L/{\rm d}z$ is the parton luminosity, defined as \cite{Dicus1989,Yang2000}
\begin{equation}
\frac{{\rm d}L}{{\rm d}z}=2z\int_{z^2}^{1}\frac{{\rm d}x}{x}g(x,\mu) g(z^2/x,\mu),
\end{equation}
where $g(x,\mu)$ and $g(z^2/x,\mu)$ are the gluon parton distribution functions.

\section{\label{sec:level3} Numerical results and conclusions}

\begin{figure}
\includegraphics{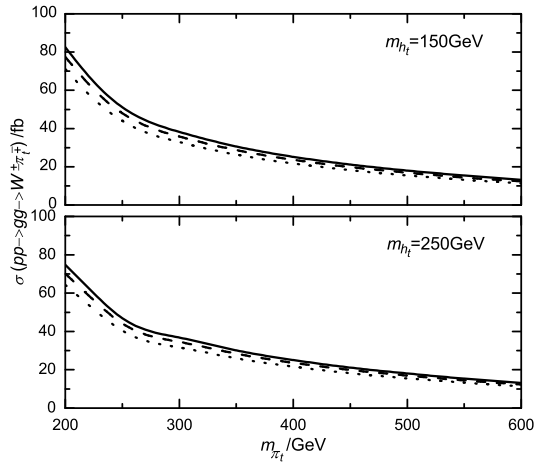}
\caption{\label{fig:eps2} The total cross section $\sigma(pp \rightarrow gg \rightarrow W^{\pm}\pi_t^{\mp})$ versus  $m_{\pi_t}$ for $\varepsilon=0.03$ (solid), 0.06 (dashed), and 0.1 (dotted).}
\end{figure}

\begin{figure}
\includegraphics{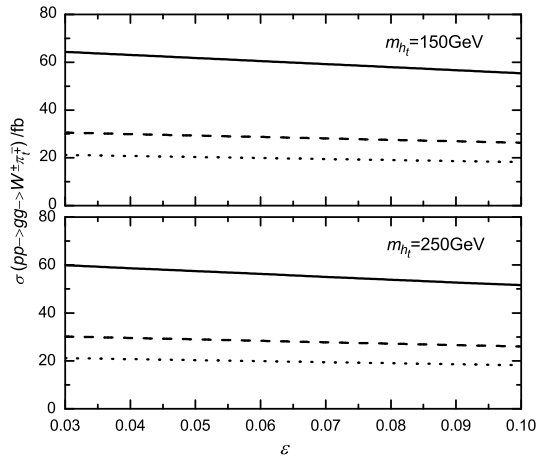}
\caption{\label{fig:eps3} The curve of $\sigma(pp \rightarrow gg \rightarrow W^{\pm}\pi_t^{\mp})$  vs. $\varepsilon$ for $m_{\pi_t}=225\ {\rm GeV}$(solid), 350 GeV (dashed), and 450 GeV
(dotted).}
\end{figure}

\begin{figure}
\includegraphics{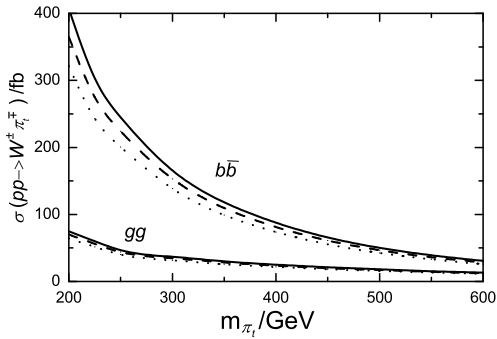}
\caption{\label{fig:eps4} The total cross section of $pp \rightarrow W^{\pm}\pi_t^{\mp}$ versus $m_{\pi_t}$ with $m_{h_t}=250\ {\rm GeV}$ for $\varepsilon=0.03$ (solid), $0.06$ (dashed) and $0.1$ (dotted).}
\end{figure}

We are now in a position to explore the phenomenological implications of our results.  The SM input parameters for our numerical analysis are $G_F=1.16639 \times 10^{-5}\ {\rm GeV}^{-2}$, $m_W=80.399\ {\rm GeV}$, $m_Z=91.1876\ {\rm GeV}$, $m_t=172.0\ {\rm GeV}$, and $m_b=4.19\ {\rm GeV}$ \cite{Nakamura2010}. We use
LoopTools \cite{Hahn1999} and the CTEQ6M parton distribution function \cite{Pumplin2006} with $\mu=\sqrt{s}/2$. The parameter $\varepsilon$ and the masses of top pion $\pi^0_t, \pi_t^{\pm}$ and top Higgs $h^0_t$ are all model-dependent \cite{Hill1995}, we select them as free parameters, and take
\begin{eqnarray}
0.03\leq \varepsilon \leq 0.1,\ 200\ {\rm GeV}\leq m_{\pi_t} \leq 600\ {\rm GeV},
\end{eqnarray}
and $m_{h_t}= 150,\ 250\ {\rm GeV}$ to estimate the total cross section of $W^{\pm}\pi_t^{\mp}$ associated production at the LHC. We sum over the final states $W^+\pi^-_t$ and $W^-\pi^+_t$ considering their symmetry. The final numerical results are summarized in Figs.
\ref{fig:eps2}-\ref{fig:eps4}.

In Fig. \ref{fig:eps2}, the total cross section $\sigma(pp \rightarrow gg \rightarrow W^{\pm}\pi_t^{\mp})$ as a function of $m_{\pi_t}$ for $m_{h_t}=150,\ 250\ {\rm GeV}$ at the LHC with $L=100\ {\rm fb}^{-1}$ is given, in which the solid lines, the dashed lines, and the dotted lines denote, respectively, the cases
of $\varepsilon=0.03, 0.06$, and $0.1$. From this diagram, we can see that (i) the total cross section decreases quickly as $m_{\pi_t}$ increase, changes the values from $77.5\ {\rm fb}$ to $12.3\ {\rm fb}$ with the range of $m_{\pi_t}$, $200 \sim 600\ {\rm GeV}$ for $\varepsilon=0.06$ and $m_{h_t}=150\ {\rm GeV}$, and from $70.2\ {\rm fb}$ to $12.3\ {\rm fb}$ for $\varepsilon=0.06$ and $m_{h_t}=250\ {\rm GeV}$, respectively; (ii) the cross section is sensitive to $\varepsilon$ and $m_{h_t}$ when $m_{\pi_t}$ is small, but this sensitivity will disappear for a rather heavy top pion; and (iii) when $m_{\pi_t}=225\ {\rm GeV}$, the cross section of $W^{\pm}\pi_t^{\mp}$ associated production via $gg$ fusion roughly $60\ {\rm fb}$, and is rather large.

Figure \ref{fig:eps3} gives the plots of the fully integrated cross section  via gg fusion versus $\varepsilon$ for $m_{\pi_t}=225,\ 350,$ and $450\ {\rm GeV}$. We can observe that (i) the cross section is not sensitive to $\varepsilon$, and only decreases by $13.8\% \sim 14.2\%$ in the range of $0.03 \leq \varepsilon \leq 0.1$ for $m_{h_t}=150\ {\rm GeV}$; and (ii) the case of $m_{h_t}=250\ {\rm GeV}$ is almost the same as that of $m_{h_t}=150\ {\rm GeV}$.

We know that there are mainly two parton subprocesses that contribute to the hadronic cross section $pp \rightarrow W^{\pm}\pi_t^{\mp}$: the $b\bar{b}$ annihilation and the $gg$ fusion. In order to look at the contributions from these two subprocess, we take $m_{h_t}=250$ and $\varepsilon=0.03$, $0.06$ and $0.1$ as an example and plot their total cross sections in Fig. \ref{fig:eps4}. We find from this figure, (i) when $m_{\pi_t}$ takes a small value, the total
cross section $\sigma (pp \rightarrow W^{\pm}\pi_{t}^{\mp})$ via $b\bar{b}$ annihilation can reach several hundred femtobarns, however, the case of $gg$ fusion is only a few dozen femtobarns, and it is evident that the former is far larger than the latter; (ii) all the total cross sections from these two subprocess are only from one dozen to two dozens femtobarns for a large value of $m_{\pi_t}$. In fact, the total cross section of $W^{\pm}\pi_t^{\mp}$ associated production at the LHC should be the sum over these two parton subprocesses.

As is known, at the LHC, the integrated luminosity is expected to reach $L=100\ {\rm fb}^{-1}$ per year. This shows that a cross section of $1\ {\rm fb}$ could translate into about 60 detectable $W^{\pm}H^{\mp}$ events per year \cite{Dicus1989,Nakamura2010}. Looking at Fig. \ref{fig:eps4}, we thus conclude that, if
$m_{\pi_t}=225\ {\rm GeV}$ and $m_{h_t}=250\ {\rm GeV}$, depending on $\varepsilon$, one should be able to collect an annual total between $1.77 \times 10^4$ and $2.18 \times 10^4$ events. So the $W^{\pm}\pi_t^{\mp}$ signal should be clearly visible at the LHC unless $m_{\pi_t}$ is very large.

In conclusion, we have calculated the technicolor corrections to the cross section for $W^{\pm}\pi_t^{\mp}$ associated production via $gg$ fusion at the CERN's LHC in the topcolor assisted technicolor model. We find that, the total cross section of $pp \rightarrow gg \rightarrow W^{\pm}\pi_t^{\mp}$, is several dozen femtobarns with reasonable values of the parameters, and the total cross section of $pp \rightarrow W^{\pm}\pi_t^{\mp}$ can reach a few hundred femtobarns when we consider the sum of the contributions of these two parton subprocesses $gg \rightarrow W^{\pm}\pi_t^{\mp}$ and $b\bar{b} \rightarrow W^{\pm}\pi_t^{\mp}$. Thus, it is so large that the signal of the charged top pion should be clearly visible at the LHC.

\vspace{-3mm}

\section*{ACKNOWLEDGMENTS}

This project was supported in part by the National Natural Science Foundation  of China under Grant Nos. 10975047 and 10979008; the Natural Science Foundation of Henan Province under Grant Nos. 092300410205 and 102300410210; and the Scientific Research Foundation of Nanyang Normal University under Grant ZX2011008.

\end{document}